\documentclass[aps,prl,twocolumn,groupaddress,amssymb,amsmath,nofootinbib]{revtex4}
\usepackage[dvipsnames]{xcolor}
\usepackage{graphicx}
\usepackage{epstopdf}
\usepackage{hyperref} 

\usepackage[caption=false]{subfig}

\begin{document}

\preprint{}

\title{Emergence of small numbers in complex systems \\and the origin of the electroweak scale}

\author{Radovan Derm\' \i\v sek}

\affiliation{Physics Department, Indiana University, Bloomington, IN 47405, USA}
\affiliation{Department of Physics and Astronomy
and Center for Theoretical Physics, Seoul National University, Seoul 151-747, Korea}

\email[]{dermisek@indiana.edu}

        %\thanks{}
%\altaffiliation {}

\date{\today}
%\date{November 10, 2016}

\begin{abstract}

In sufficiently complex models with many parameters that are unknown or undetermined from first principles, a small  coupling or mass can naturally arise even if it  is not protected by a symmetry or a result of some dynamics. For the naturalness criterion, we advocate specifying all the model parameters with one significant figure. This automatically avoids outcomes of a given observable that would require special highly-tuned choices for parameters. The obtained range of outcomes is an attribute  of a given model resulting from its specific characteristics and complexity.  
%Within the obtained range, any outcome  is indistinguishable with respect to the way model parameters are specified.
%In this way we  will  demonstrate that, when specific characteristics and complexity of  supersymmetric extensions of the SM are taken into account,
%the electroweak scale up to 3 orders of magnitude below superpartner masses naturally occurs. 
Using these criteria in the minimal supersymmetric model, we demonstrate that the electroweak scale up to 3 orders of magnitude below superpartner masses naturally occurs.
%, without fine tuning of model parameters, and it is not probabilistically disfavored compared to any other possibility.

\end{abstract}


\maketitle

%\section{ \label{sec:}}

%\subsection{}

%\subsubsection{}

%\subsection{Introduction}

%\subsubsection{Experimental results}

\noindent
{\bf Introduction.}
In particle physics,  it is usually argued that  a parameter, written in units of maximal energy at which a given model is valid, can naturally be significantly smaller than unity if it is protected by a symmetry from quantum corrections due to various interactions~\cite{Wilson:1970ag, 'tHooft:1980xb}. %Since there is no symmetry in the standard model (SM) protecting the Higgs boson mass  from radiative corrections, this immediately implies the existence of new physics not far above the electroweak EW scale~\cite{Gildener:1976ai, Weinberg:1978ym, Susskind:1978ms, Veltman:1980mj}.
When applied to the Higgs boson mass, which is related to the electroweak (EW) scale, this immediately implies the existence of a new physics not far above the EW scale, since there is no symmetry in the standard model (SM) protecting the Higgs mass from radiative corrections~\cite{Gildener:1976ai, Weinberg:1978ym, Susskind:1978ms, Veltman:1980mj}. The naturalness problem related to the Higgs boson mass is one of the  most studied problems in particle physics and the main motivation for physics beyond the standard model.

Realizing that supersymmetry (SUSY) can provide the needed symmetry for any scalar mass~\cite{Witten:1981nf},  together with the intriguing picture of unification of strong and electroweak forces~\cite{Dimopoulos:1981yj} and the possibility to build simple realistic models~\cite{Dimopoulos:1981zb} made supersymmetric extensions of the SM  the most promising candidates for new physics. In these models, the EW scale is a result of quantum corrections and  is calculable. These quantum corrections are comparable to masses of superpartners of SM particles  and thus a generic prediction of these models, based on the naturalness argument, is that superpartners are at or very near the EW scale~\cite{Martin:1997ns}. 

So far negative results of searches  for superpartners, with some limits exceeding an order of magnitude above the EW scale, cast a shadow on this framework (similar concerns apply to other proposals for physics beyond the SM).
Moreover, the measured value of the Higgs boson mass, taking aside naturalness considerations, indirectly points to superpartners two orders of magnitude above the EW scale~\cite{Draper:2013oza}. Since the relevant parameters are masses squared, this would mean that the EW scale corresponds to a number which is 4 orders of magnitude smaller than expected. It is commonly argued that such an outcome, apparently contradicting naturalness principle,  would require fine tuning of model parameters at the level of 1 part in $10^4$ and thus it is very unlikely~\cite{Martin:1997ns}.

In this letter we argue that  in sufficiently complex models with many parameters that are unknown or undetermined from first principles, a small  coupling or mass can naturally arise even if it  is not protected by a symmetry or a result of some dynamics.   Rather than relying on various probabilistic arguments or measures of sensitivity, the naturalness criterion adopted in this letter advocates specifying all the model parameters with one significant figure. This automatically avoids outcomes of a given observable that would require special highly-tuned choices for parameters. The obtained range of outcomes is an attribute  of a given model resulting from its specific characteristics and complexity.  Within that range any outcome  is indistinguishable with respect to the way model parameters are specified and thus is considered natural. Using these criteria, we will  demonstrate that in the minimal supersymmetric model 
the electroweak scale up to 3 orders of magnitude below superpartner masses naturally occurs.

Besides complexity, an important characteristic of the EW symmetry breaking in supersymmetric models is that the EW scale is typically a result of  the cancellation of comparable contributions. 
In a toy model that closely mimics features of electroweak symmetry breaking, we will show that in such situations, even if all the model parameters are specified with just one significant figure, the outcomes  several orders of magnitude smaller compared to dominant contributions arise with probabilities comparable to the most likely outcome.

\begin{figure}[t]
\subfloat[]{
\includegraphics[width=1.6in]{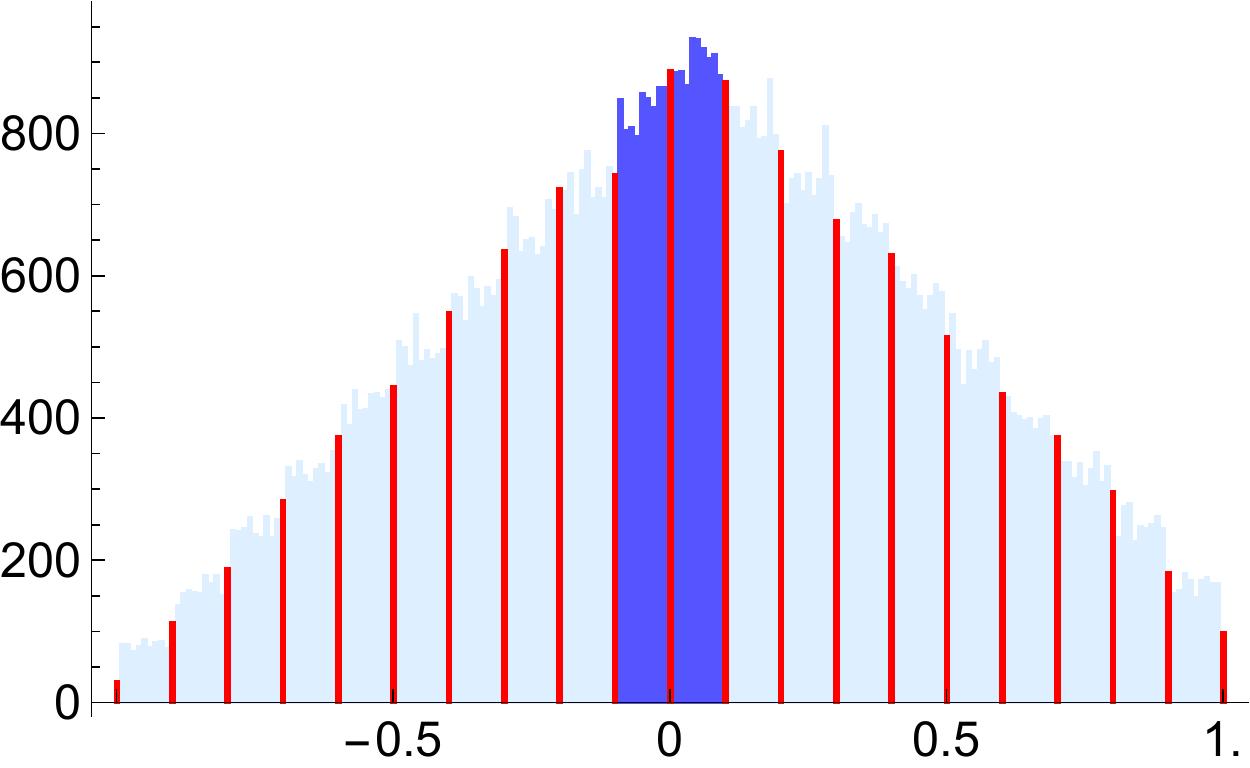}
}
\subfloat[]{
\includegraphics[width=1.6in]{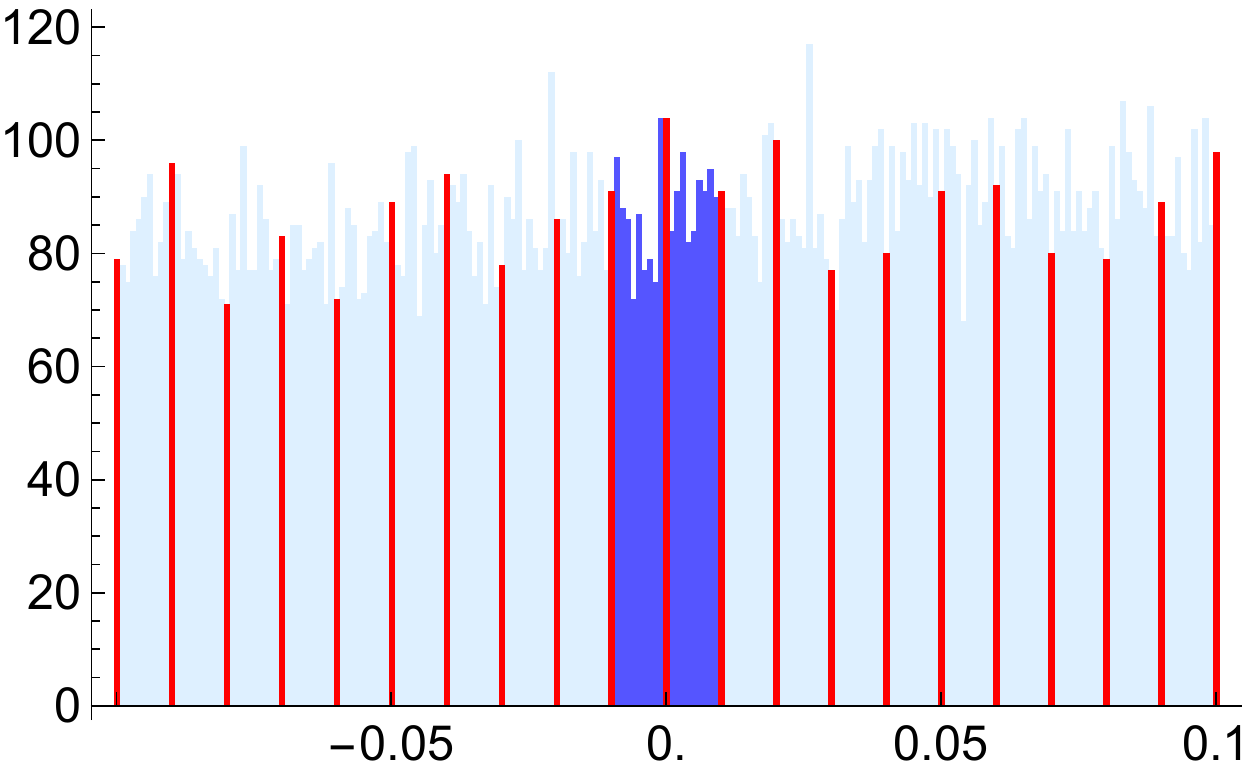}
}
\\ 
\vspace{0.3cm}
\subfloat[]{
\includegraphics[width=1.6in]{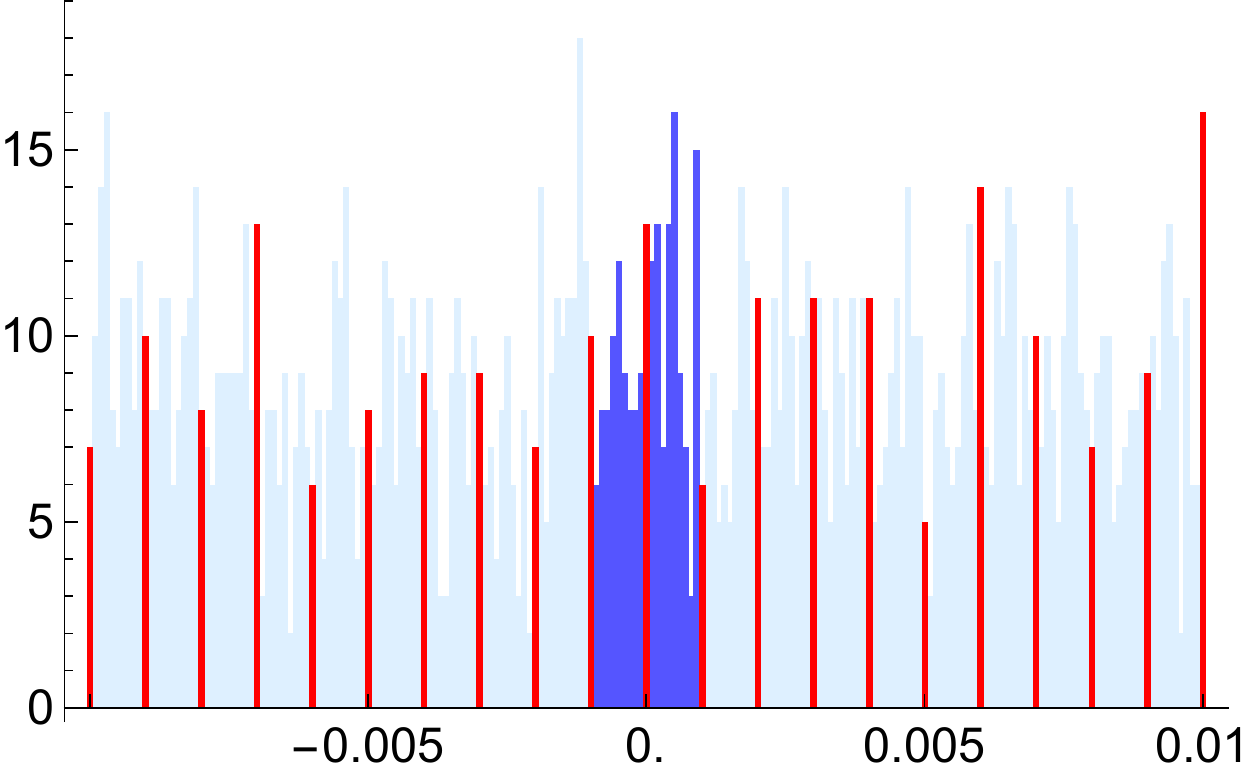}
}
\subfloat[]{
\includegraphics[width=1.6in]{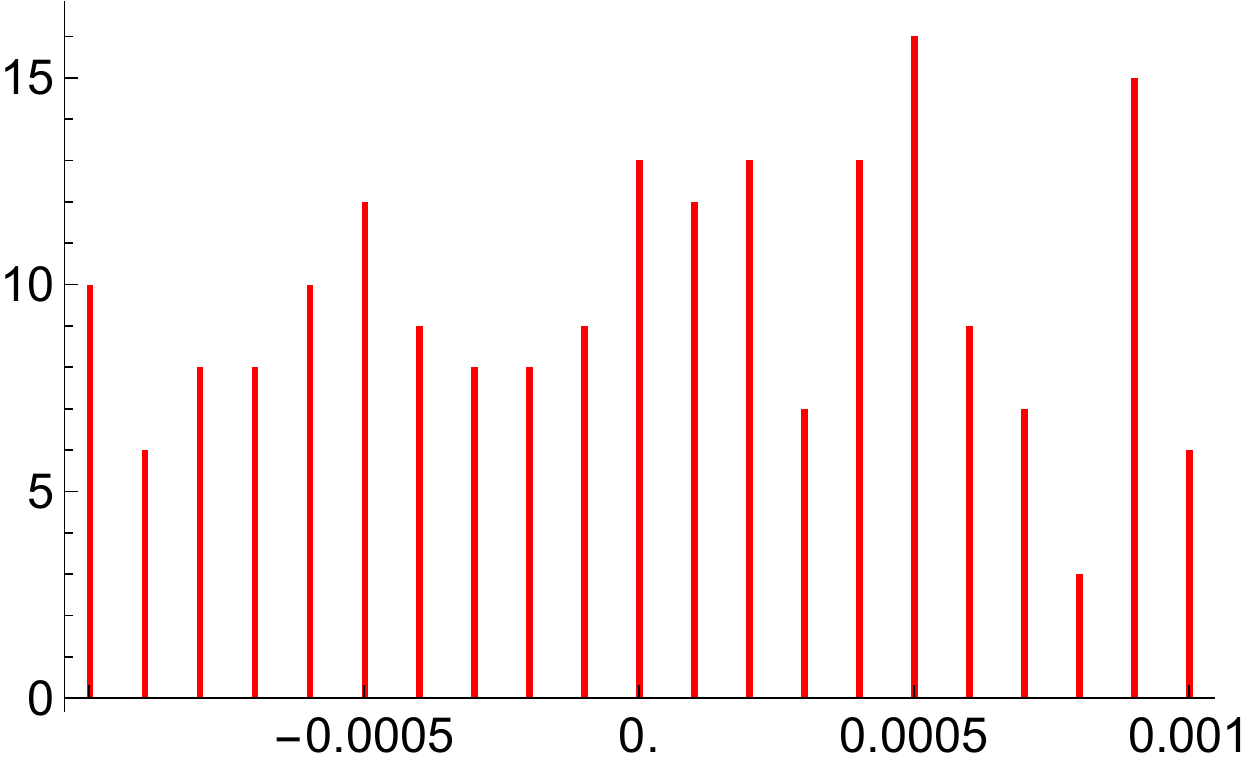}
}
\caption{Distribution of $X = R[1]-R[1]+R[0.1]+ R[0.01]+R[0.001]$, where $R[x]$ is a randomly generated number from interval $[0.5x,1.5x]$ keeping only the first significant figure specifying the departure from $x$. For example,  R[0.1] randomly generates numbers $0.1+\{0,\pm0.01, \pm0.02, ..., \pm 0.05\}$. The bin size of  histograms is 0.01 (a), 0.001 (b), 0.0001 (c) and 0.00001 (d). The highlighted region near zero in plot (a) is shown in plot (b) and so on. The highlighted lines (red) are the outcomes of $R[1]-R[1]$ (a),  $R[1]-R[1]+R[0.1]$ (b), $R[1]-R[1]+R[0.1]+ R[0.01]$ (c) and the complete $X$ in (d). } 
\label{fig:H}
\end{figure}

\vspace{0.2cm}
\noindent
{\bf Toy model.}
Let us consider an observable $X$, which is, up to much smaller contributions from other parameters in the model, given by the difference of two parameters $A$ and $B$ of comparable size:
\begin{equation}
X \simeq A-B.
\end{equation}
We can always choose appropriate units so that $A$ and $B$ are random real numbers close to 1 and, in order to be specific, let us allow them to vary by 50\% in both directions, between 0.5 and 1.5. The distribution of $X$ is then an almost symmetric triangle peaked at 0, similar to that in Fig.~\ref{fig:H}(a) (which is slightly shifted to the right due to perturbations). For unequal intervals of $A$ and $B$ or their central values the distribution could be less peaked, more broad and shifted to the left or right. Nevertheless, as far as there is a non-negligible overlap in intervals over which $A$ and $B$ are allowed to vary, an outcome in the vicinity of zero will have a comparable probability to the most probable outcome.

If $A$ and $B$ were integers in overlapping ranges, there would be no problem associated with $X$ being a small number. It would be just one out of many possible comparably likely outcomes. Being integers supplies a natural bin size in which the outcomes are plotted and compared. For real numbers, an arbitrarily small  outcome of X is possible. However it would require specifying A and B with many significant figures and having almost identical values. This, without any symmetry or dynamical reason, would seem like a huge coincidence and it is the essence of the fine tuning problem.

For example, if we specify the departure of $A$ and $B$ from 1 by one significant figure, the outcomes of $X$ will be distributed in intervals of 0.1, indicated by highlighted lines in Fig.~\ref{fig:H}(a). 
%{\color{brown} In this case the smaller the outcome the more likely it is, but that is just a consequence of equal overlapping intervals. For unequal intervals of $A$ and $B$ or their central values the distribution would be less peaked, more broad and shifted to the left or right. Nevertheless, as far as there is a non-negligible overlap in intervals over which A and B are allowed to vary, the outcome 0.1 will have comparable probability to the most probable outcome.} 
A mismatch of the central values of $A$ and $B$ would shift the 0 outcome, but since the outcomes appear in 0.1 intervals, an $X$ as small as 0.1 with either sign will naturally appear with  probability comparable to the most likely outcome.  A significantly smaller $X$ would however require specifying  departures of $A$ and $B$ from 1 by more than one significant figure.

However, in many physical systems there are more than two parameters contributing to a given observable. Let us consider a parameter $C$ contributing to $X$ at  an order of magnitude smaller level than $A$ and $B$. The exact value of $C$ is not important since it only shifts the $A-B$ distribution slightly to the left or right and thus we can simply assume it is equal to 0.1.  Let us again allow the parameter $C$ vary in the interval  $\pm 50\%$ around the central value, similarly to what we allowed for parameters $A$ and $B$, and specify  the departure from the central value with one significant figure.  Adding $C$ is then, up to a constant shift, equivalent to adding random numbers: $0.1+\{0, \pm0.01,\pm0.02, ...,\pm 0.05\}$. Let us define such a set of random numbers as  $R[0.1]$. In such a notation, our $A$ and $B$ specified with one significant figure are $R[1]$.  Varying $C$ in steps corresponding to specifying it with one significant figure results in emergence of new  outcomes of $X$ in intervals of 0.01, or $0.1 C$. These new outcomes are indicated by highlighted lines in Fig.~\ref{fig:H}(b).  

If there are more parameters contributing at smaller levels, even if  specified by just one significant figure, new outcomes will appear in intervals  $\simeq 0.1 $ of the corresponding parameter. The distribution of $X$ resulting from adding  three such perturbations, $ R[0.1]+R[0.01]+R[0.001]$, is shown in different intervals in Fig.~\ref{fig:H}(a)-(d). In this specific example, outcomes as small as $10^{-4}$ appear and they are still about 10 times more likely than outcomes of order 1. No parameter needed to be specified with more than one significant figure, no tuning was required. 

In order to obtain,  without fine tuning, outcomes of X orders of magnitude smaller than contributions from $A$ or $B$, it is necessary that a given model is sufficiently complex with at least several more parameters contributing to it.   More importantly, contributions of these parameters have to be distributed in a way that the space of possible outcomes is continuously covered without any gaps. If, for example, there was a new parameter  contributing to $X$ at $10^{-8}$ level, it would not change our previous  conclusions.  This new contribution would only split the existing outcomes into new outcomes  tightly clustered around those shown in Fig.~\ref{fig:H}(d). 
Because of the arbitrary shift from all the parameters of the model, from digits that were not specified, the naturally occurring outcomes would still not be smaller than $10^{-4}$.  It is the largest gap in the distribution that determines the smallest naturally occurring outcome from the minimally specified model parameters. It is easy to see that our toy model in Fig.~\ref{fig:H} is optimal for achieving the smallest outcome.

The obtained range of outcomes is a characteristic of a given model. While probabilities of individual outcomes depend on intervals over which parameters are allowed to vary and prior distributions of model parameters, the range itself is independent on these assumptions. For example, it is very easy to change the central values of the parameters in the toy model in Fig.~\ref{fig:H} so that the distribution peaks at any other value. Nevertheless, as far as intervals of $A$ and $B$ overlap, outcomes of $X$ as small as $10^{-4}$ are inevitable. However, no matter what assumptions we make, outcomes of $X$ smaller than $10^{-4}$ cannot be guaranteed without specifying model parameters with more than one digit. Such outcomes would disappear after an arbitrary shift and thus  they are purely accidental.

Armed with the intuition from the toy model we can find the range of hierarchy between  the electroweak scale and  the scale of superpartners that corresponds to specifying parameters in the minimal sypersymmetric model  with one significant figure. 
As in our toy model, the electroweak scale is a result of the cancellation of comparable numbers. In addition,  the complexity of these models is enormous, with more than 100 parameters contributing to the determination of the EW scale.

\vspace{0.2cm}
\noindent
{\bf EW symmetry breaking.}
In the standard model, EW symmetry breaking is the result of  a negative quadratic term and a positive quartic term in the Higgs potential. The minimum of the potential, corresponding to the vacuum expectation value of the Higgs field, is away from zero and determines the scale of electroweak interactions and masses of $Z$, $W$ and Higgs bosons. 

One of the most attractive features of supersymmetric extensions of the SM is the fact that the negative mass squared term, triggering EW symmetry breaking, is obtained by radiative corrections in vast ranges of parameters of various models. The relevant mass squared parameter is a combination of the supersymmetric Higgs mass parameter,  $\mu$,  and the 
soft SUSY breaking mass squared parameter, $m_{H_u}^2$, 
\begin{equation}
| \mu |^2 + m_{H_u,0}^2 + \Delta m_{H_u}^2 ,
\label{eq:mass2}
\end{equation}
where we intentionally split the $m_{H_u}^2$ into the value generated by a SUSY breaking scenario at a given scale, $m_{H_u,0}^2$, and the radiative correction from the renormalization group (RG) running to the electroweak scale, $\Delta m_{H_u}^2$. The $|\mu|^2$ part could  also  be split in a similar way, but radiative corrections to this term are not dramatic and thus we use the EW scale value directly. The negative of the combination of  parameters in Eq.~(\ref{eq:mass2})  is directly related to the $Z$ boson mass or the Higgs mass and thus it should  be close to $(100 \;{\rm GeV})^2$~\cite{Martin:1997ns}.

\begin{figure}[t]
\subfloat[$M_{SUSY} = m_0$, $M_{1/2} = 0$.]{
\includegraphics[width=1.6in]{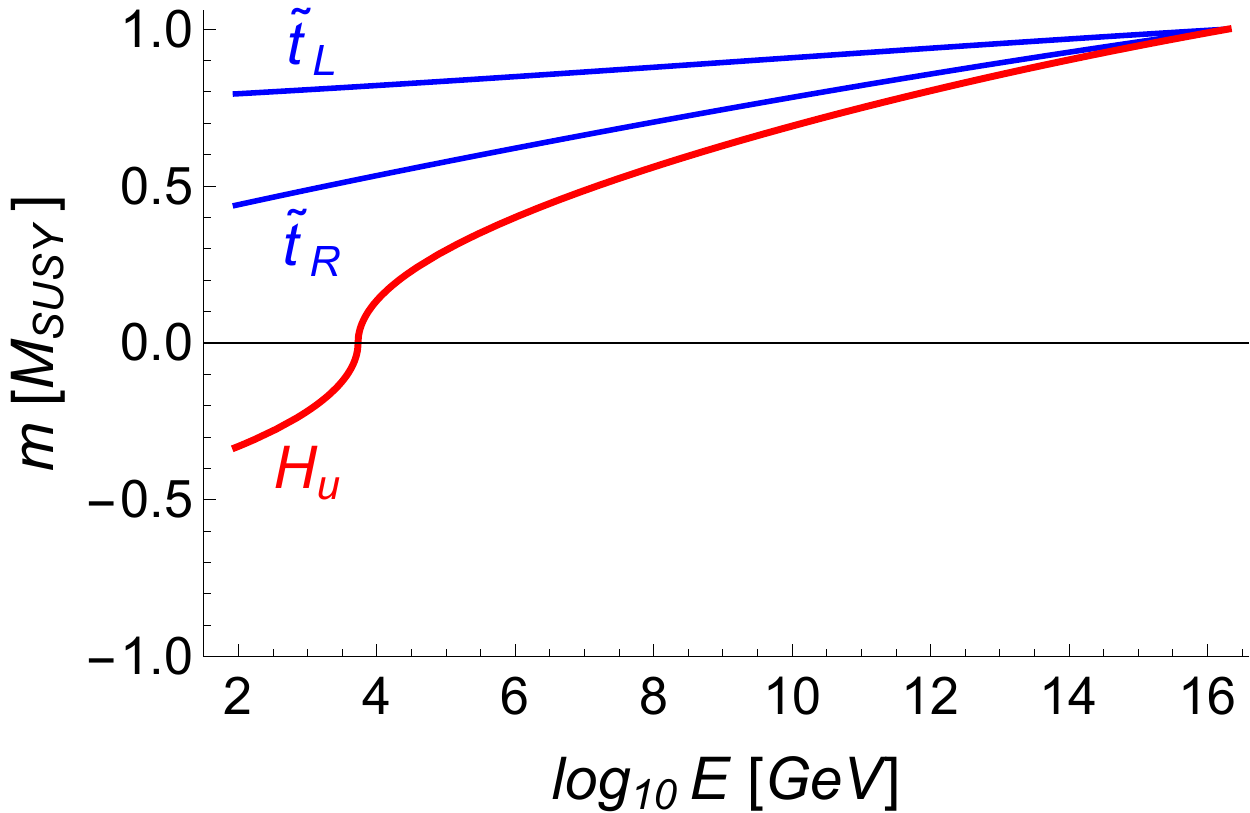}
}
\subfloat[$M_{SUSY} = M_{1/2}$, $m_0 = 0$.]{
\includegraphics[width=1.6in]{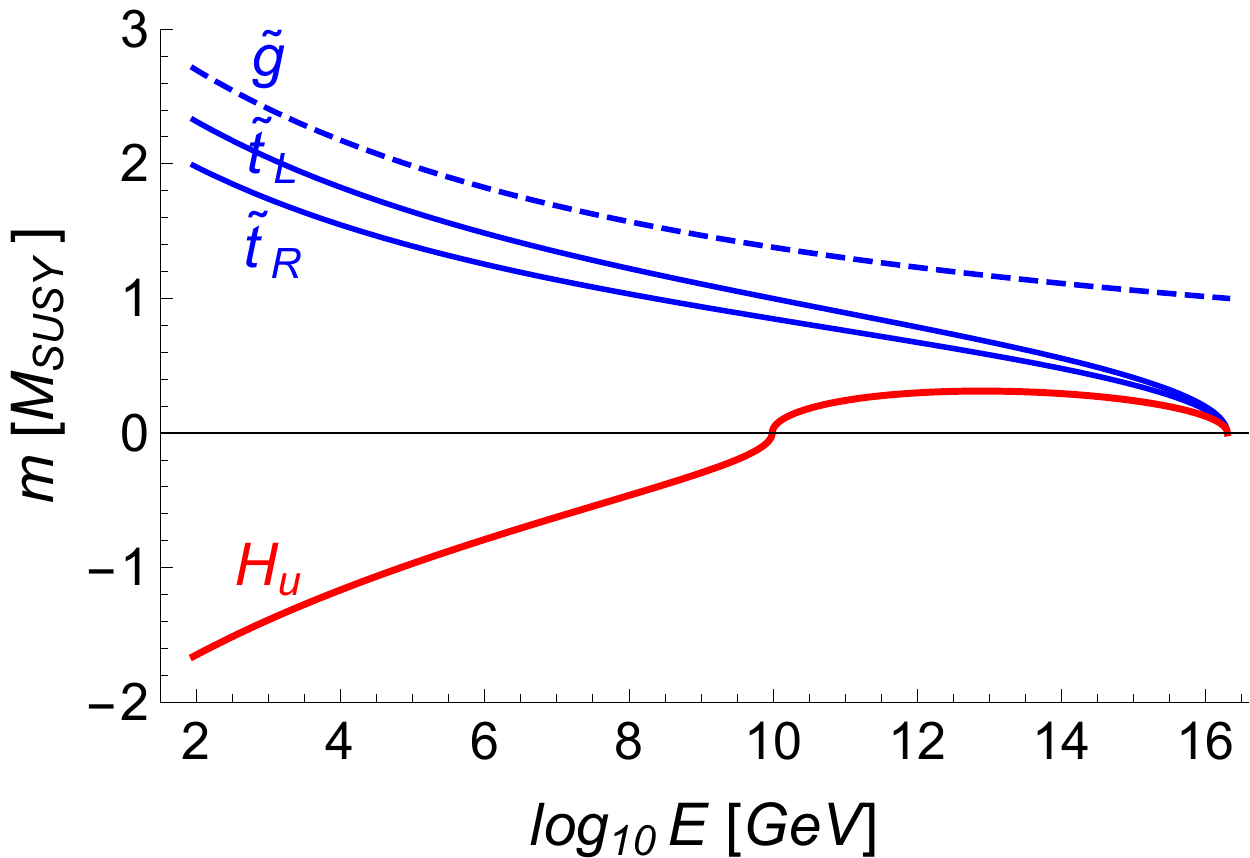}
}\\
%\vspace{0.3cm}
\subfloat[$M_{SUSY} = m_0=M_{1/2}$.]{
\includegraphics[width=1.6in]{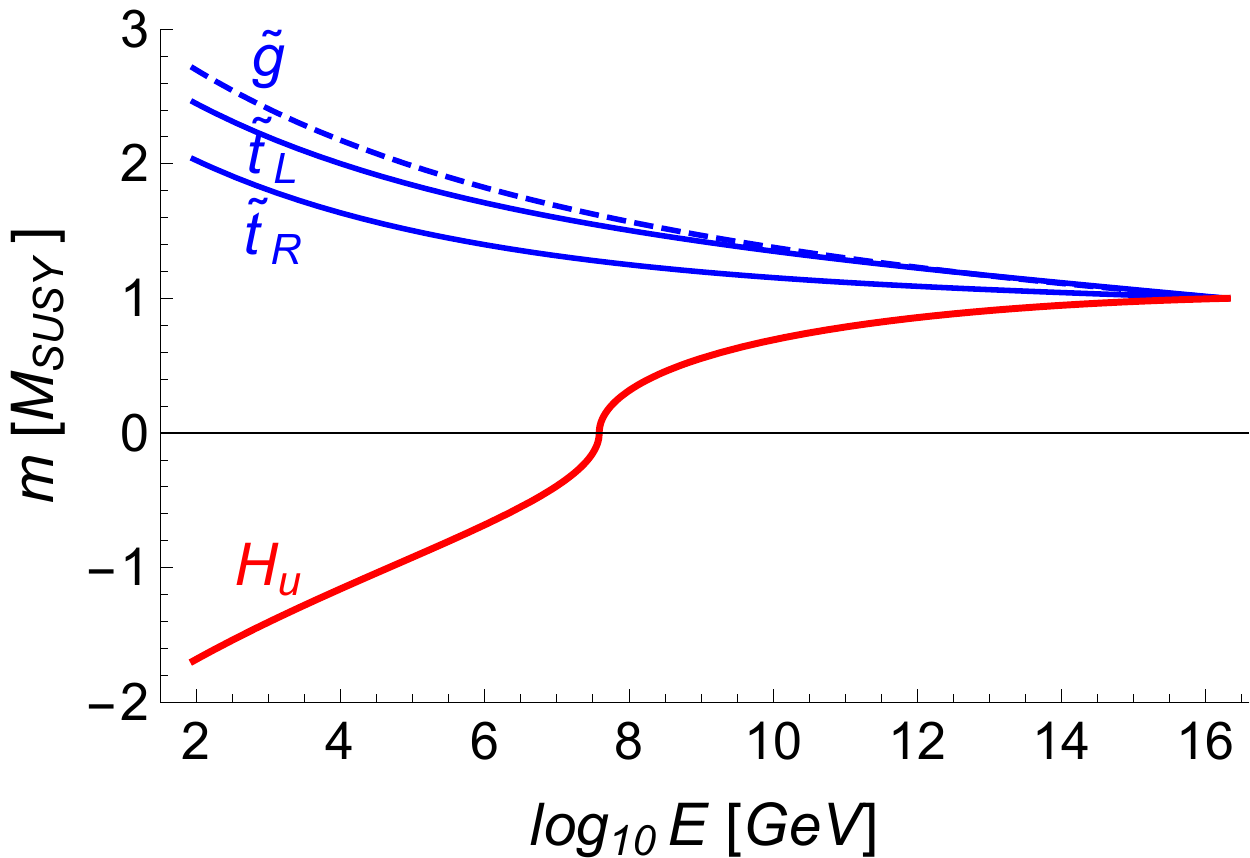}
}
\hspace{-0.4cm}
\subfloat[]{
\includegraphics[width=1.7in]{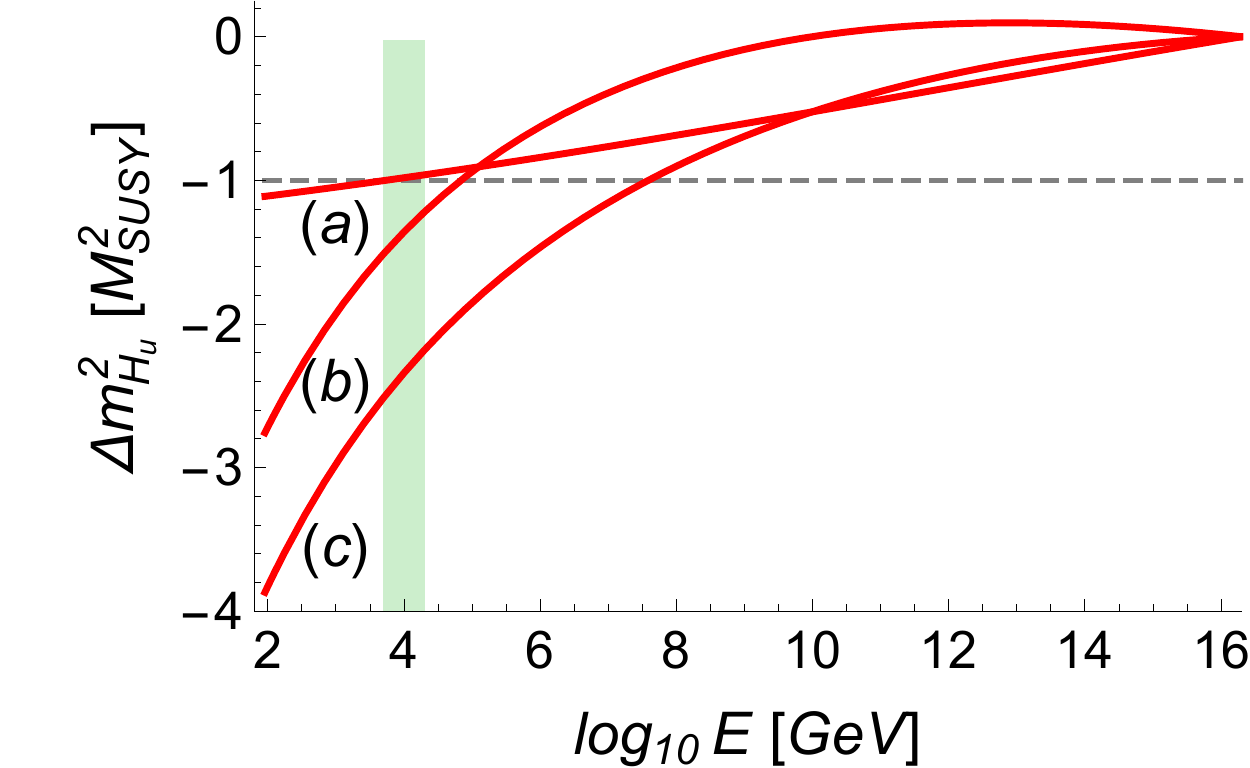}
}
\caption{RG evolution of the most relevant parameters contributing to the electroweak symmetry breaking.  In (a), (b) and (c) the mass parameters of the gluino, $\tilde g$, left and right stops, $\tilde t_{L,R}$, and $H_u$, plotted as $m_{H_u}^2/|m_{H_u}^2|^{1/2}$ to account for $m_{H_u}^2$ turning negative at low energies, are plotted for three different boundary conditions in units of $M_{SUSY}$. In (d), the radiative correction, $\Delta m_{H_u}^2$, is plotted for the three cases, (a)--(c).
The shaded region, $5 - 20$ TeV, represents the  range of stop masses which can result in the measured value of the Higgs boson mass without significant additional contributions. 
}
\label{fig:RG}
\end{figure}

The power of radiative corrections is illustrated in Fig.~\ref{fig:RG} (a)--(c) where the RG evolution of  $m_{H_u}$ is shown together with the most contributing  other parameters, the stop masses and the gluino mass for three representative cases. Although all superpartner masses can be free parameters, simple scenarios with just few parameters are commonly assumed. Here we assume a universal scalar mass, $m_0$, a universal gaugino mass, $M_{1/2}$, and consider three cases: $M_{1/2} = 0$, $m_0 = 0$ and $M_{1/2} = m_0$. We also define $M_{SUSY} = \max(M_{1/2}, m_0)$ and the RG evolution in Fig.~\ref{fig:RG} is plotted in this unit starting  at the grand unification scale (it is another simple and common assumption although it is  not necessary for our discussion). In all three cases, and thus for any case in between,  the $m_{H_u}^2$ parameter turns negative.

Remarkable feature of the RG flow is that $\Delta m_{H_u}^2$ eventually reaches $- M_{SUSY}^2$ but it doesn't run to much larger negative values. More specifically, for the case of $m_0$ domination,  the $\Delta m_{H_u}^2 \simeq - M_{SUSY}^2$ after about 12 orders of magnitude of the RG flow. Similarly, in the case of gaugino domination, the same situation is achieved after 11 orders of magnitude of the RG flow and finally for $M_{1/2} = m_0$ we need about 9 orders of magnitude. In further RG evolution $\Delta m_{H_u}^2$ does not exceed $- 1.2 M_{SUSY}^2$, $-3 M_{SUSY}^2$, and   $-4 M_{SUSY}^2$ respectively, even if evolved all the way to the EW scale which can be seen in Fig.~\ref{fig:RG} (d). Note, however, that stop masses no longer contribute to $\Delta m_{H_u}^2$ at energies below their masses. For stop masses which can result in the measured value of the Higgs boson mass without significant  additional contributions, $5 - 20$ TeV, indicated by shaded region in Fig.~\ref{fig:RG} (d), the $\Delta m_{H_u}^2$ ranges between $-1 M_{SUSY}^2$ and $-2.5 M_{SUSY}^2$.

The value of the $\mu$ term in Eq.~(\ref{eq:mass2}) or its origin, although subject to intense research and usually the center of attention in naturalness considerations,   is not crucial for our discussion. We assume $\mu \lesssim M_{SUSY}$ and sufficiently small not to prevent electroweak symmetry breaking. 
With  $| \mu |^2 + m_{H_u,0}^2 \simeq  M_{SUSY}^2$, either because $m_{H_u,0}\simeq M_{SUSY}$ or $\mu \simeq M_{SUSY}$ or both, we can see that the  negative of the mass squared combination in Eq.~(\ref{eq:mass2}) that sets the EW scale squared, $M_{EW}^2$, is the result of cancellation of two comparable contributions as in our toy model: $X = - M_{EW}^2$, $A = | \mu |^2 + m_{H_u,0}^2  \simeq  M_{SUSY}^2$ and $B= -\Delta m_{H_u}^2  \simeq  M_{SUSY}^2$.

Although written in a simple way, the mass squared parameter in Eq.~(\ref{eq:mass2}) depends on every single soft SUSY breaking scalar and gaugino mass, every gauge coupling and every Yukawa coupling in the model because they all contribute in the RG evolution to $\Delta m_{H_u}^2$. Assuming all soft masses comparable (not necessarily equal) of order $M_{SUSY}$ at the grand unification scale, from  two loop RG equations~\cite{Martin:1993zk} we find that the approximate individual contributions to $\Delta m_{H_u}^2$  in units of  $M_{SUSY}^2$ are as follows: -2.6 from gluino, -0.3 from each scalar top and $H_u$, 0.2 from the $SU(2)$ gaugino, 0.05 from the $U(1)_Y$ gaugino, $-4\times 10^{-3}$ from the trace of masses squared of all $SU(2)$ doublet scalars,  $-4\times 10^{-4}$ from the trace of masses squared of all scalars carrying hypercharge,  $10^{-3}-10^{-5}$ from each scalar bottom and $H_d$  (for the ratio of vacuum expectation values of two Higgs doublets, $\tan\beta$, in the range $\tan\beta = 50 - 5$), and $-4\times 10^{-6}$ from each scalar charm.
After scalar charm contributions, there is a gap of about 5 orders of magnitude. Each scalar up contributes at $3\times 10^{-11}$ and the contribution from each scalar strange is about $10^{-8}$ of the contribution from scalar bottom.

In the list above, we did not include contributions from neutrino Yukawa couplings and scalar neutrinos, since these couplings are not known and are highly model dependent, and thus also from scalar charged leptons since contributions of these depend on the neutrino sector. In addition, we omitted the hypercharge weighted trace of all scalar masses squared which does not contribute if  scalar masses are equal. Furthermore, we assumed diagonal Yukawa and scalar mass matrices, and neglected contributions from soft trilinear couplings that, if flavor diagonal, would contribute at comparable levels to scalar masses.

 % both stop masses and the gluino mass with coefficient ranging from 0.66, 0.34 to 2.6. At $0.22 M_{SUSY}^2$ level it is affected by the SU(2) gaugino mass, $M_2$, and the $U(1)_Y$ gaugino mass contributes at about $0.05 M_{SUSY}^2$. The hypercharge weighted trace of all scalar masses squared contributes at about $0.025 M_{SUSY}^2$ (for equal masses it would be 0). Each of scalar bottoms and $H_d$ contributes at $10^{-3}-10^{-5}$ for $\tan\beta = 50 - 5$. Or combined contribution from terms involving bottom Yukawa coupling is 6 times that. Trace of masses of SU(2) doublet scalars contributes at $4\times 10^{-3}$  and similar trace over scalars carrying hypercharge contributes at $4\times 10^{-4}$. Individual scalar charm $4\times 10^{-6}$ and all terms proportional to charm Yukawa 3 times that. Individual scalar up $3\times 10^{-11}$ and all terms proportional to charm Yukawa 3 times that and also scalar strange is $10^{-8}$ compared to sbottom.

We have found that  contributions of various parameters to $\Delta m_{H_u}^2$ in first 6 orders of magnitude below $M_{SUSY}^2$ are very dense, with 13 parameters contributing, followed by a gap to below $10^{-10}$ level. As in the toy model, having many new parameters contributing  in 6 orders of magnitude below $M_{SUSY}^2$ without gaps means that,  specifying with one significant figure the way they are varied  around central values, the outcomes for the electroweak scale squared as small as $10^{-6}\; M_{SUSY}^2$, or $M_{EW} \simeq10^{-3}\; M_{SUSY}$,  occur  with probabilities comparable to the most likely outcome.

As already discussed for the toy model, probabilities of individual outcomes do depend on the choice of intervals and prior distributions of model parameters, but the range of outcomes, $10^{-3} M_{SUSY} \lesssim M_{EW} \lesssim M_{SUSY}$, is a characteristic of the minimal supersymmetric model with all the mass parameters of the same order at sufficiently high scale and specified with one digit. 
The range of hierarchy between the EW scale and $M_{SUSY}$ characteristic to constrained versions of the MSSM will be explored in Ref.~\cite{MSSM}. 
%However, having 13 parameters contributing is significantly more than in the toy model and thus the conclusions would be comparable for large variations in the chosen procedure.

\vspace{0.2cm}
\noindent
{\bf Discussion.}
Conclusions drawn in this letter are in a sharp contrast with commonly accepted views. In our analysis,  the EW scale two orders of magnitude below superpartner masses, as suggested by the Higgs boson mass,   is well within the range of comparably likely outcomes, $10^{-3} M_{SUSY} \lesssim M_{EW} \lesssim M_{SUSY}$,  that do not need specifying any of the parameters with more than one significant figure. On the other hand, it is commonly  argued that the EW scale resulting from 10 TeV superpartners, requires fine tuning of model parameters  at the level of 1 part in $10^4$. This is usually estimated using various probabilistic arguments or  Barbieri-Giudice measure, $|\partial \ln M_{EW}^2/\partial \ln p_i|$, where $p_i$ are model parameters and the role of $M_{EW}$ is played by the Z boson mass, the Higgs mass or the vacuum expectation value in various versions~\cite{Barbieri:1987fn}\cite{Martin:1997ns}. 

This quantity, being a derivative, clearly represents the sensitivity of the EW scale to model parameters.  However, as we showed, this sensitivity does not necessarily indicate the need for special choices of parameters. In our toy model, the outcome $X \simeq 10^{-4}$ corresponds to  $|\partial \ln X/\partial \ln A| \simeq 10^4$, possibly suggesting fine tuning needed at the level of 1 part in $10^4$ just like for the EW symmetry breaking with 10 TeV superpartners. If there were just two model parameters, $A$ and $B$, this would indeed be the level of fine tuning required: for any $A$, the $B$ would have to be specified with 4 significant figures. However, with more parameters contributing at smaller levels without significant gaps, none of the parameters needed to be specified with more than 1 significant figure. The Barbieri-Giudice measure does not take into account the complexity of a given model and indicates the same for a model with two parameters as for a model with many parameters.

 The naturalness criterion adopted in this letter automatically avoids outcomes of a given observable that would require special highly-tuned choices for parameters. The obtained range of outcomes is an attribute  of a given model resulting from its specific characteristics and complexity.  Within that range any outcome, even the smallest one,  is indistinguishable with respect to the way model parameters are specified.  If the obtained range is large, the small probability of an individual outcome is a reflection of the complexity of a given model rather than a sign of unnaturalness.

%%%%%%%%%%%%%%%%%%%%%%%%%%%%%%%%%%%%%%%%%%%%%%%%%%%%%%%%
%\acknowledgements
%%%%%%%%%%%%%%%%%%%%%%%%%%%%%%%%%%%%%%%%%%%%%%%%%%%%%%%%

\vspace{0.2cm}
\noindent
{\it Acknowledgments:} This work was supported in part by the U.S. Department of Energy under grant number {DE}-SC0010120
and by  the Ministry of Science, ICT and Planning (MSIP), South Korea, through the Brain Pool Program.

%%%%%%%%%%%%%%%%%%%%%%%%%%%%%%%%%%%%%%%%%%%%%%%%%%%%%%%%

%%%%%%%%%%%%%%%%%%%%%%%%%%%%%%%%%%%%%%%%%%%%%%%%%%%%%%%%%

%%%%%%%%%%%%%%%%%%%%%%%%%%%%%%%%%%%%%%%%%%%%%%%%%%%%%%%%%
\end{document}